\PassOptionsToPackage{sort&compress}{natbib} 
\documentclass[preprint,12pt]{elsarticle}

\usepackage{amssymb}
\usepackage{amsmath}
\usepackage{color}
\usepackage{url}

\usepackage{lineno}

\journal{Nuclear Instruments and Methods A}

\newcommand\code[1]{\texttt{#1}}

\newcommand\added[1]{#1}

\begin{document}
\emergencystretch 3em

\begin{frontmatter}



\title{An integrated online radioassay data storage and analytics tool for nEXO}


\author[Alabama]{R.H.M.~Tsang}
\author[Alabama]{A.~Piepke}
\author[McGill]{S.~Al Kharusi}
\author[Stanford]{E.~Angelico}
\author[PNNL]{I.J.~Arnquist}
\author[Drexel]{A.~Atencio}
\author[Carleton]{I.~Badhrees \fnref{fn6}} \fntext[fn6]{Permanent at: King Abdulaziz City for Science and Technology, Riyadh, Saudia Arabia}
\author[UMass]{J.~Bane}
\author[ITEP]{V.~Belov}
\author[LLNL]{E.P.~Bernard}
% ...
\address[Alabama]{Department of Physics and Astronomy, University of Alabama, Tuscaloosa, AL 35405, USA}
\address[McGill]{Physics Department, McGill University, Montr\'eal, Qu\'ebec H3A 2T8, Canada}
\address[Stanford]{Physics Department, Stanford University, Stanford, CA 94305, USA}
\address[PNNL]{Pacific Northwest National Laboratory, Richland, WA 99352, USA}
\address[Drexel]{Department of Physics, Drexel University, Philadelphia, PA 19104, USA}
\address[Carleton]{Department of Physics, Carleton University, Ottawa, Ontario, K1S 5B6, Canada}
\address[UMass]{Amherst Center for Fundamental Interactions and Physics Department, University of Massachusetts, Amherst, MA 01003, USA}
\address[ITEP]{National Research Center ``Kurchatov Institute'', Moscow, 123182, Russia}
\address[LLNL]{Lawrence Livermore National Laboratory, Livermore, CA 94550, USA}
% ...

\begin{abstract}
Large-scale low-background detectors are 
increasingly used in rare-event searches as experimental collaborations push for enhanced sensitivity.
However, building such detectors, in practice, 
creates an abundance of radioassay data especially during the conceptual phase of an experiment
when hundreds of materials are screened for radiopurity.
A tool is needed to manage and make use of the radioassay screening data to quantitatively assess detector design options.
We have developed a Materials Database Application for the nEXO experiment to serve this purpose.
This paper describes this database \added{application}, explains how it functions, and discusses how it streamlines the design of the experiment.
\end{abstract}


\end{frontmatter}


\section{Introduction}

Neutrinoless double beta decay ($0\nu\beta\beta$) is a hypothetical nuclear process that violates lepton number conservation.
Its detection would prove that neutrinos are Majorana particles \cite{schechtervalle}. 
The next-generation Enriched Xenon Observatory (nEXO) \cite{pcdr}, a successor to the EXO-200 \cite{exo200} experiment,
aims to observe $0\nu\beta\beta$ using a liquid xenon (LXe) time projection chamber (TPC) 
using approximately 5 tonnes of xenon enriched to 90\% in $^{136}$Xe. 
nEXO is projected to reach a half-life sensitivity of $1.35\times10^{28}$ years at 90\% C.L. with 10 years of data taking \cite{sens2}.
To reach this sensitivity, controlling background is a major challenge.

The majority of the background comes from the detector components, directly from their intrinsic radioactivity or radon emission. 
To search for appropriately radiopure materials that also satisfy 
mechanical, cryogenic, and other constraints, hundreds of materials are screened.
The necessary radioassay campaign generates a large amount of data that is being used in the design of the nEXO detector.
Thus, there is a need for a tool to store the data, and to simplify the use and the interpretation of the data in the design process.
The nEXO Materials Database 
Application \added{(the Application)}, which retrieves information from its associated Materials Database \added{(the Database)}, is aimed to satisfy this need.
In this paper, 
we discuss the requirements for the \added{Application}, its implementation, and its performance.
\added{Discussion on the nEXO background model is beyond the scope of this paper and is described in detail in \cite{sens1,sens2}.}

\section{Detector design process and the requirements for the Application}

\begin{figure*}
\includegraphics[width=\textwidth]{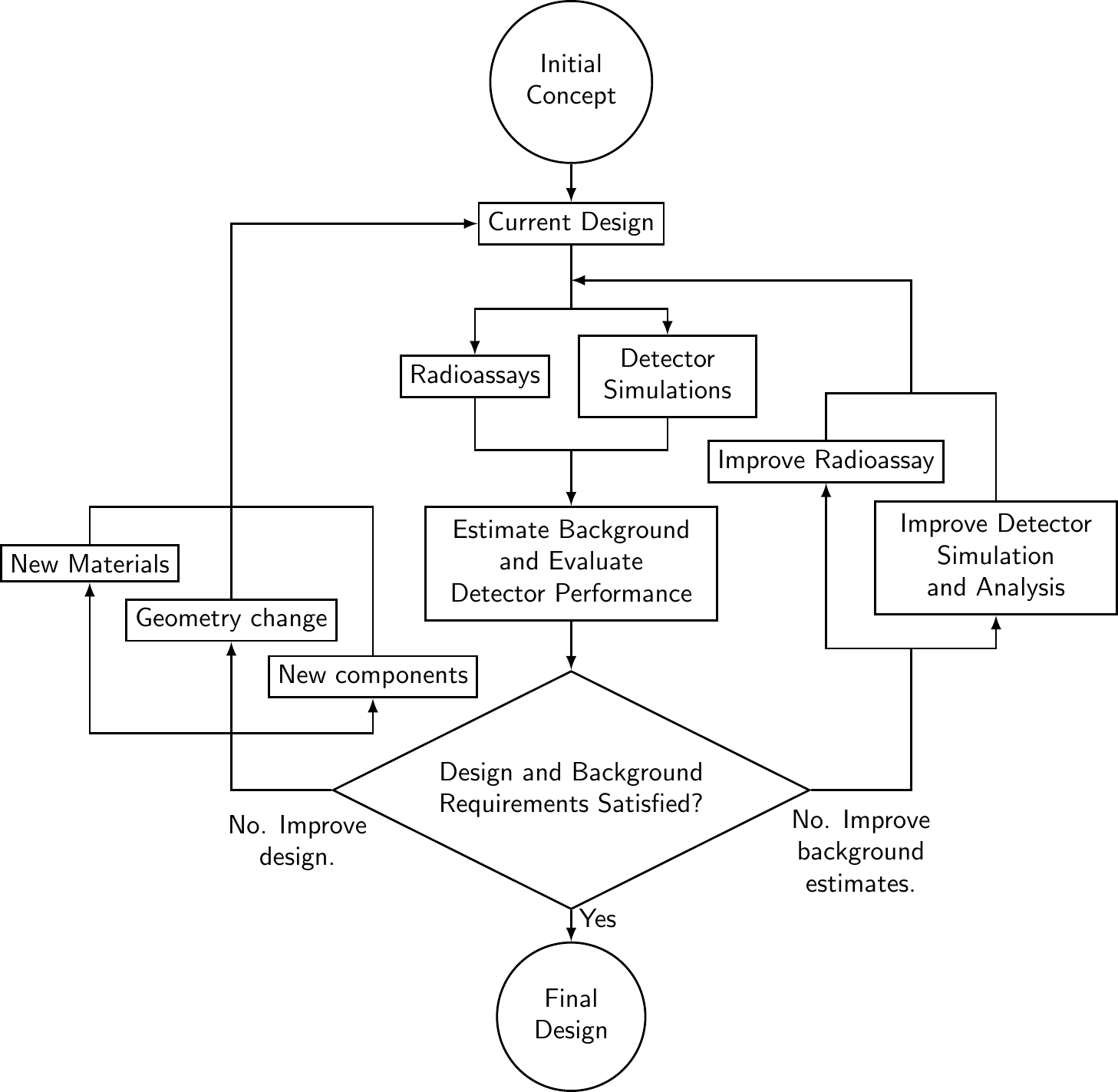} \\
\caption{Flowchart for the design process of a typical ultra-low background physics detector.}
\label{fig:designcycle}
\end{figure*}

The main purpose of the Application is to facilitate the detector design process as illustrated in Figure~\ref{fig:designcycle}.
nEXO's detector design process, like other low-background experiments, is an iterative process that begins with 
an initial concept based on experience of past experiments and physics goals,
and is then successively updated towards a feasible detector through the collective contributions of the collaboration.
For the Application, detector design consists of a list of components specified by their masses, materials, geometrical shapes, positions, etc.
It also includes a complete construction and assembly procedure to account for background events due to exposure to cosmic rays, radon, and dust.
The detector design is converted into a GEANT4 \cite{geant4a, geant4b, geant4c} model for the simulation of radioactive decays, particle transport, and interactions. 
The simulation provides the probability of each background source to produce a count in the detector as a function of energy and position -- 
we call these ``hit efficiency probability density functions (PDFs)''.

The radioactive content of all materials used in the detector design have to be carefully assessed.
Some are based on previous radioassay measurements
\cite{radsno, radborexino, radlaub, radexo200a, radgerda, radnext, radedelweiss, radpandax, radmajorana, radxenon1t, radexo200b, radlz} 
when applicable,
while some are newly measured.
In case of exposure-based backgrounds, the anticipated exposure conditions during construction and assembly are also taken into account.
Combining all of this information results in a total background rate estimate, in addition to the relative contribution of each component, 
helping to arrive at a tangible interpretation of the data.
Review of the various background contributions may reveal that
the background rates, created by some components, are higher than desirable and an alternative design is needed.
This initiates the iterative process of reducing the mass, replacing the material, or moving components to a new position in the detector setup.
New components may also be added at this stage as the engineering design of the detector is being developed.
The new design will be evaluated in the next iteration of the design process.
The iterative process continues until there is a design that satisfies all necessary requirements, both engineering design and background level.

The requirement for the Application is threefold. 
First, it needs to facilitate background rate estimation for evaluation of detector performance.
Calculating a realistic background rate estimate requires information from 
the radioassay effort, the simulation effort, and various subsystem design groups. 
Such information needs to be saved,  cataloged, and versioned so that it is easily retrievable.
Version control is especially important, for example, in the preparation of publications to be able to track updates or corrections to data made after collecting it from ``tagged'' versions.
It also needs to provide tools for data interpretation, in the form of background rate estimation.
Second, closing the loop in the flowchart, the Application needs to assist in
the investigation of potential design improvements. 
Trade studies are regularly performed to compare different design choices.
Background is an important factor in such studies in addition to other technical requirements.
The Application needs to provide tools to easily compare different material and geometry choices.
Finally, since nEXO collaborators are located around the world, permission-controlled and secure web-based access is necessary.
Accessibility also refers to connections to the Application from other databases of the collaboration. 
This means that some form of software interface between computer programs is needed.

\section{Implementation}

Radioassay data management tools have been developed by other experimental collaborations, such as, Radiopurity.org \cite{radiopurity} and Background Explorer \cite{bgexp}.
These tools are useful in their own use cases. However, they do not meet nEXO's specific needs.
The approach described here evolved out of a conceptually similar toolset used in EXO-200
which had a similar organizational structure of linked log entries and reports, data fields, and lookups for background calculations. 
Some overview of this evolution was presented at LRT2017 \cite{lrt2017}. 
Guided by previous experience, the nEXO Materials Database Application is designed to be completely web-based
with tighter integration between data fields and logging/reports. 
This is particularly important as data must be connected to realities of batch procurement, parts construction, sample preparation, cleaning, available quantities etc.,
and tight integration between data fields and written logs or reports facilitates review and consistency. 
Specifically,
a web application using a CouchDB \cite{couchdb} database as the backend storage was created.
Complex computations are supported by Flask, a Python web framework.
In the following two sections,
the Database and the Application will be described in detail
from the users' and the developer's perspectives.
\added{A list of libraries used in the Application is shown in \ref{sec:liblist}.}

\section{Functional perspective}

From the users' point of view, the Application serves four functions: 
storage of data,
calculation of background rates for data interpretation,
providing interfaces to other systems, and coordination of assay requests.

\subsection{Storage of radioassay measurement results}

\begin{figure*}
\includegraphics[width=\textwidth,trim=7 490 10 10,clip]{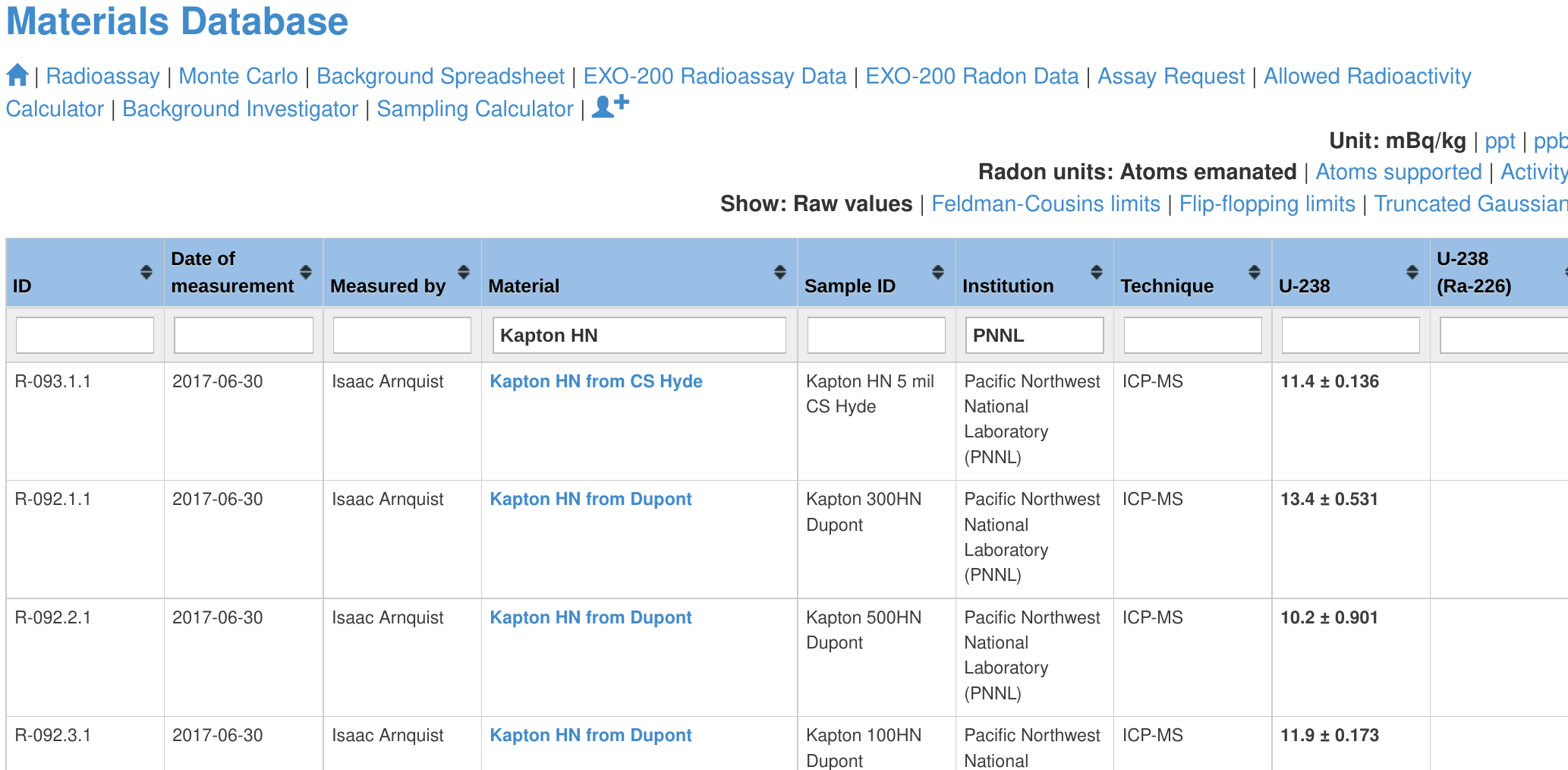}\\
\caption{A screenshot of an HTML table showing radioassay measurement results for some Kapton samples published in \cite{kapton}.
The navigation bar on top provides links to other tables and tools. 
The switches in top right corner allows users to select the unit and the kind of limits to display.
Some rows and columns are cropped out from the image due to space limitations.
}
\label{fig:bigtable}
\end{figure*}

When searching for a radiopure material, numerous measurements of a group of similar materials are often made.
For this reason, 
the Database supports a hierarchical structure for storing measurement data, see Figure~\ref{fig:schema} for details.
Each material is stored as a CouchDB document. In a document, a material may have multiple samples.
For a given sample, multiple measurements may have been performed by different radioassay techniques.
Data from a measurement may have been analyzed multiple times using different methods. 
The hierarchical document structure allows for storing all of this information in one document per material.
This structure also allows supplementary information such as,
sample descriptions, personnel involved, dates of measurements, and other metadata to be saved at the same level in the tree structure.
Another benefit of having a hierarchical structure is that 
measurement results that have been superseded by a more sensitive measurement can be marked as obsolete
while keeping them in the same document for bookkeeping purposes. 
Photos, catalog pages, and external reports can also be attached to the documents in the Database.
We found the ability to unambiguously document and identify materials or components particularly useful to avoid the duplication of analysis effort.
To reflect the structure, measurement records are identified by an ID in the format of R-xxx.y.z.t, where
xxx is the material identification number, y is the sample number, z is the measurement number, and t is the analysis number.
This is intended to be a human-readable identifier not used as an internal key in the Database.
Figure~\ref{fig:bigtable} shows a screenshot of the Application's user interface displaying a table of radioassay data.

The Materials Database serves as the primary bookkeeping tool for nEXO's radioactive background control program. 
As far as radioassay measurements are concerned, the Database stores three types of quantities:
\begin{itemize}
\item Specific activities of radioisotopes, such as $^{238}$U and $^{232}$Th, and their progeny\added{:} $^{226}$Ra and $^{210}$Pb for $^{238}$U and $^{228}$Ra for $^{232}$Th.
These are typically measured by $\gamma$-counting with high purity Ge detectors and reported in mBq/kg or mBq/cm$^2$.
\added{Also under this category are other radioisotopes as impurities in materials, cosmogenic isotopes, such as $^{60}$Co, and $\alpha$-emitters that can produce $^{137}$Xe via $(\alpha,n)$ reactions.}
\item Concentrations of elements of interest, such as U and Th, normalized by mass, (e.g. ng/g, pg/g) or area.
These are mostly measured by 
 inductively coupled plasma mass spectrometry (ICP-MS) and neutron activation analysis (NAA), and sometimes by
glow discharge mass spectrometry (GD-MS) and accelerator mass spectrometry (AMS).
\item Radon emanation rates, reported in atoms per day or atoms per day per cm$^2$,
usually measured by electrostatic counting (ESC) or liquid scintillation counting. The Database accepts emanation rates of both $^{222}$Rn and $^{220}$Rn.
\end{itemize}
While these are different physical quantities, they are closely related.
The specific activity of a radioisotope can be converted to a concentration of the element when secular equilibrium and natural isotopic abundance are assumed.
The Application provides convenient tools for converting between concentration units and specific activity units.
Although radon emanation ultimately comes from disintegration of $^{238}$U, and $^{232}$Th, 
the fraction of which that emanates in the form of radon is still an active research area \cite{rnemanfrac}.
Hence, no such tools are available in the Application for converting between radon and $^{238}$U or $^{232}$Th.

Measurement results are typically reported in the form of a central value with an associated one-standard deviation (1-$\sigma$) error that is symmetric around the central value.
The Application provides tools to convert the reported measurement into a 90\% C.L. limit (when appropriate), using the Feldman-Cousins method \cite{fc} or 
the so-called ``flip-flopping'' method, also described in \cite{fc}, where the limit is formed by adding to the central value a certain multiple of the uncertainty.
As a way to simplify comparison between measurements, a truncated Gaussian (TG) distribution can be defined by the reported central value and error.
\added{
The TG distribution, whose probability density function $f(x|\mu,\sigma)$ is defined below, is a Gaussian distribution truncated at zero, with central value $\mu$ and error $\sigma$.
\begin{equation}
f(x|\mu,\sigma) = \begin{cases} 
          0 & x \le 0 \\
          \frac{A}{\sigma\sqrt{2\pi}} \cdot e^{-\frac{(x-\mu)^2}{2\sigma^2}} & x > 0 
       \end{cases}
\end{equation}
where $A$ is the normalization constant.
When only an upper limit is given,
it will be treated as if $\mu = 0$ and $\sigma = s$ where $s$ is chosen so that the area under the TG distribution from 0 to the given limit is consistent with the stated confidence level.
With this unified treatment, limits can be combined with measurement results using ordinary error propagation as verified by toy Monte Carlo.}
The TG mean allows to retain some information about the mean instead of just reporting a limit, and does so in a way that disregards negative possibilities.
Likewise it allows calculation of limits using 90\% of total probability of non-negative values. 
Published data from the EXO-200 materials assay effort \cite{exorad1, exorad2} has been imported into the Database, allowing their inclusion in background calculations.

\subsection{Storage of simulation results}

Background rate calculations require hit efficiencies of every component, which are in the form of ROOT files produced 
by GEANT4-based simulation code custom-built for nEXO.
The simulation results are grouped into documents \added{and are stored as attachments}.
Each document contains the simulation results for all radioisotopes in a detector component determined in a simulation campaign, in addition to the \added{metadata of} the Monte Carlo calculation.
\added{Neutron yields calculated are also stored in this document when relevant.}
\added{The schema of the Monte Carlo document is shown in Figure \ref{fig:schema}.}
The web interface generates different projections of the hit efficiency PDFs and displays them as interactive histograms.
\added{For example, it can display the hit efficiency for signal-like events as a function of energy.}

\subsection{Background rate calculation}

In addition to being a repository of information, another important purpose of the Application is to serve as a background rate calculator to enable the interpretation of the data it contains.
To perform a full background calculation, the user first creates a detector specification document which contains a table of all detector components. 
Each row of the table contains the mass of the component, the document IDs for the material the component is made of (its R-identification number) and
the detector simulation \added{document that contains its hit efficiencies}.
The detector specification document can either be manually entered or modified from an existing specification document using the
\code{Background Investigator}\footnote[1]{The names of custom-made software components of the Application are in \code{monospace} font.}
tool. 
The user can first make a copy of an existing background model, then make edits to the copy of the model.
Once the table is complete, 
the Application collates all the information and allows the user to download a spreadsheet summarizing all background contributions.
The background spreadsheet is a self-contained calculator for detector background rates. 
In the current version, it has 18 sheets including the title page.
10 sheets contain radioassay and simulation data from the Database to be used as input. 
They can be modified by the user to quickly investigate the effects of materials with different radioactivity or mass.
The main output of the calculations, background rates from every component due to different radioisotopes, are reported in the next 6 sheets.
These sheets display the numerical background rates along with summary charts showing breakdowns of background contribution by material, component, and isotope.
A final sheet details the background budget allocation for each engineering subsystem in the nEXO Work Breakdown Structure in comparison with the background model at hand.

\begin{figure*}
\includegraphics[width=\textwidth,trim=5 440 55 20,clip]{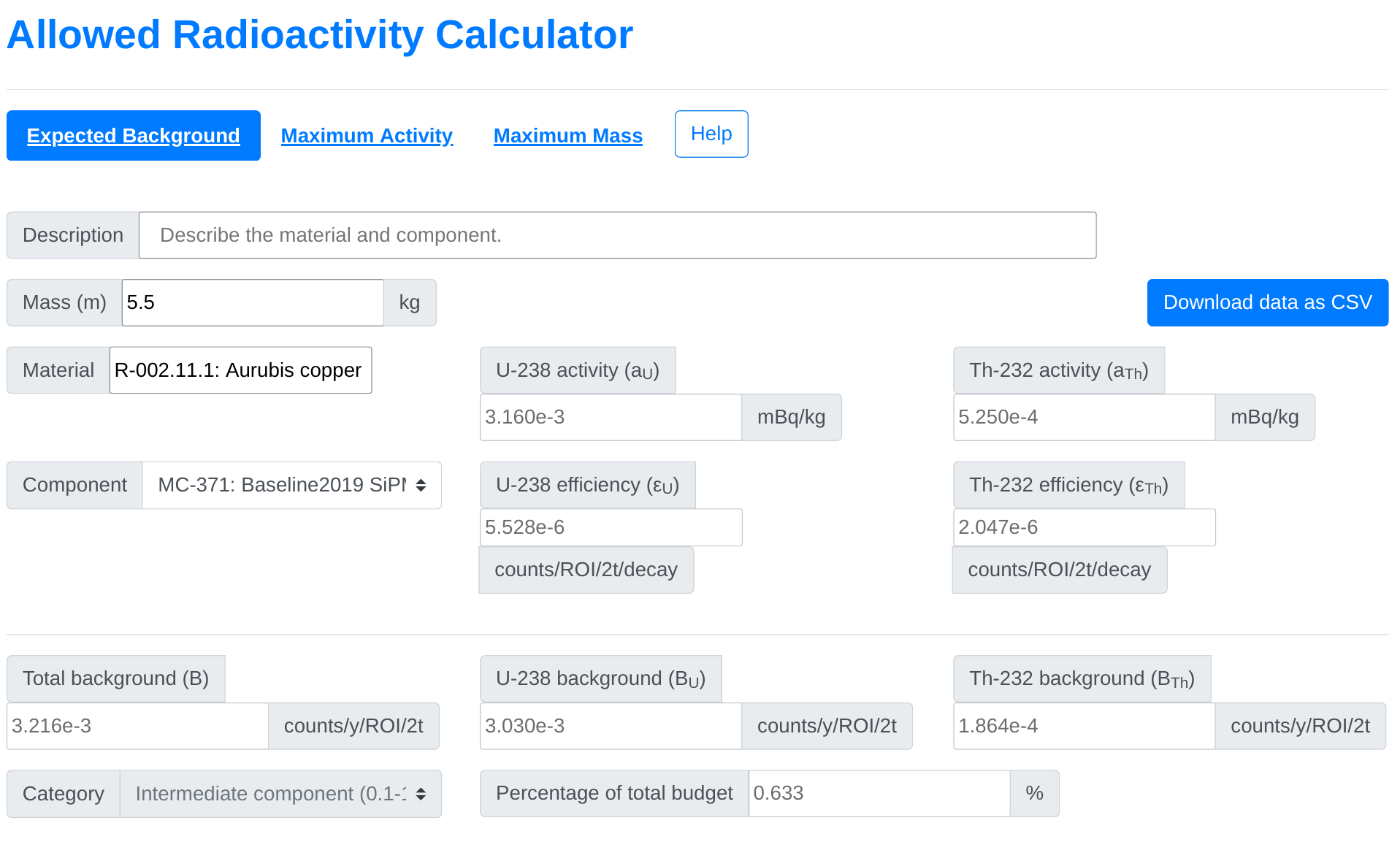} \\
\caption{A screenshot of the \code{Allowed Radioactivity Calculator}.}
\label{fig:arc}
\end{figure*}

The \code{Allowed Radioactivity Calculator}, a lightweight online background calculator, is also provided by the Application. 
Its primary usage is for 
calculating the allowed specific activities \added{of $^{238}$U and $^{232}$Th of} a component. 
\added{It considers U and Th decay chains only because simulations for other isotopes are only performed as needed and therefore generally unavailable.}
It takes as input the mass, the hit efficiencies for the U and Th decay chains of the component,
and the background allocated to the component.
The background allocation plan is formulated by radioactive background control, simulation, and subsystem experts based on current expectation of its background contribution.
The calculator is also capable of performing calculations in two other ways.
It can calculate the maximum allowed mass of the component given the allowed background, the hit efficiencies and the specific activities of the material.
It can also calculate the background contribution of one component, given the mass, the hit efficiencies, and the U and Th decay chain specific activities of the material.
It draws radioassay and simulation data from the Database, making it a convenient tool for investigating material and location choices.
The calculation result can be downloaded as a comma-separated-value (CSV) file for record.
Figure~\ref{fig:arc} shows a screenshot of the calculator.

\subsection{Coordination of Assay Requests}
As seen in Figure~\ref{fig:designcycle}, the assay of materials requires coordination between different project design subsystems.
The Application features a table where assay requests can be entered. An automated email will be sent to relevant personnel responsible for radioassay measurements. 
This single channel of communication reduces the amount of email exchanges and possible confusion. 

\section{Software structure}

\begin{figure*}
\includegraphics[width=\textwidth]{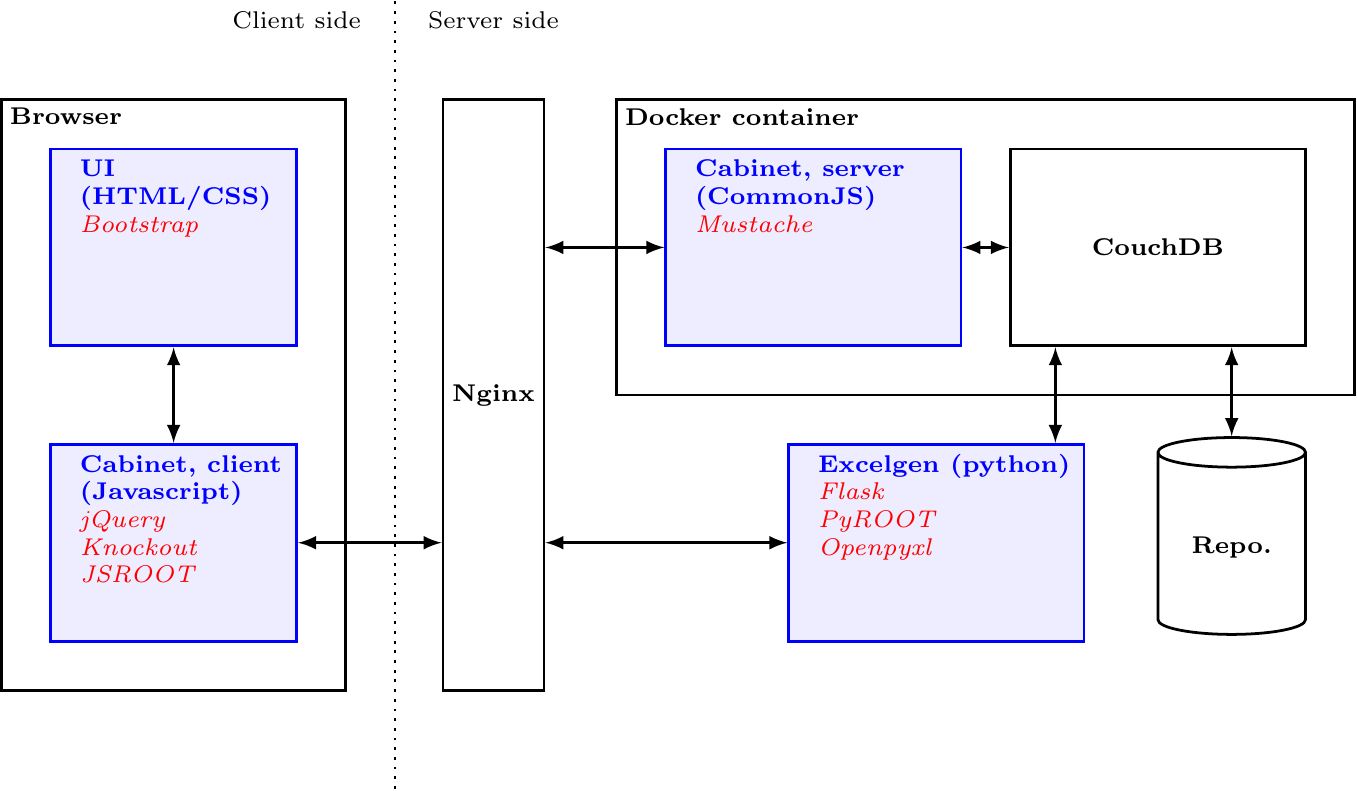} \\
\caption{Schematic of the software configuration. Components of the Application are \added{shaded} in blue. The libraries and external software that they make use of are in red \added{italics}.}
\label{fig:schematic}
\end{figure*}

As illustrated in Figure~\ref{fig:schematic},
the Application is structured like a typical web based application -- a frontend that runs on the client browser, and a backend that runs on the server which also hosts \added{the CouchDB} database. 
Being a non-relational document store,
CouchDB stores data in the form of JavaScript Object Notation (JSON) \cite{json} documents, allowing more flexibility in document structure than relational databases.
JSON is a standard file format that 
stores data as key-value pairs and arrays
in plain human-readable text.
Key-value pairs and arrays can be nested, 
For example, to implement a multi-level hierarchical structure in \added{relational databases such as} MySQL, one table will be needed for each level, 
whereas in CouchDB, the structure can be represented in JSON so that only a single repository is needed.
Combined with its Hypertext Transfer Protocol (HTTP) Application Programming Interface (API) and built-in JavaScript interpreter,
it is possible to store a web application\added{, both the frontend and the backend,} as a document in CouchDB. 
These kinds of web applications are called CouchApps.
Database commands, such as, create, read, update, and delete are sent as a Uniform Resource Identifiers (URI).
Parameters of the operations are encoded in the URI or sent along with a request as ``POST'' data. 
This feature simplifies the task of interfacing the database with other software used by the collaboration.

The CouchApp developed for the nEXO Materials Database is called \code{Cabinet}.
\code{Cabinet} makes extensive use of jQuery and Bootstrap in developing the frontend user interface.
\added{
JQuery is a JavaScript library, primarily for HTML manipulation and event handling.
Due to its high level of abstraction, the created code is often much shorter and hence easier to maintain.
}
\added{
Bootstrap is a frontend toolkit that provides default styles and layouts for responsive browser display.
It simplifies developers' work on the aesthetic aspect of the design of a user interface.
}
\added{The use of jQuery and Bootstrap} has reduced the time needed to develop a user-friendly frontend.

Although JavaScript is among the most popular languages on the web, it has some limitations.
JavaScript has limited support for ROOT \cite{root}, an analysis framework that is widely used in nuclear and particle physics.
Therefore, an alternative is needed for complex computations that involve ROOT objects.
For the case of the Application, 
ROOT is needed for reading detector simulation results that are saved as ROOT objects.
The solution is to create a custom Python web application, called \code{Excelgen}, based on the Flask web framework.
\code{Excelgen}, being a Python application, can read ROOT files using PyROOT and perform background rate calculations.
It also has access to Python packages for creating complex spreadsheets.
Details on the components of the Application are discussed in the following subsections.

\subsection{Database component}

\begin{figure*}
\includegraphics[width=\textwidth]{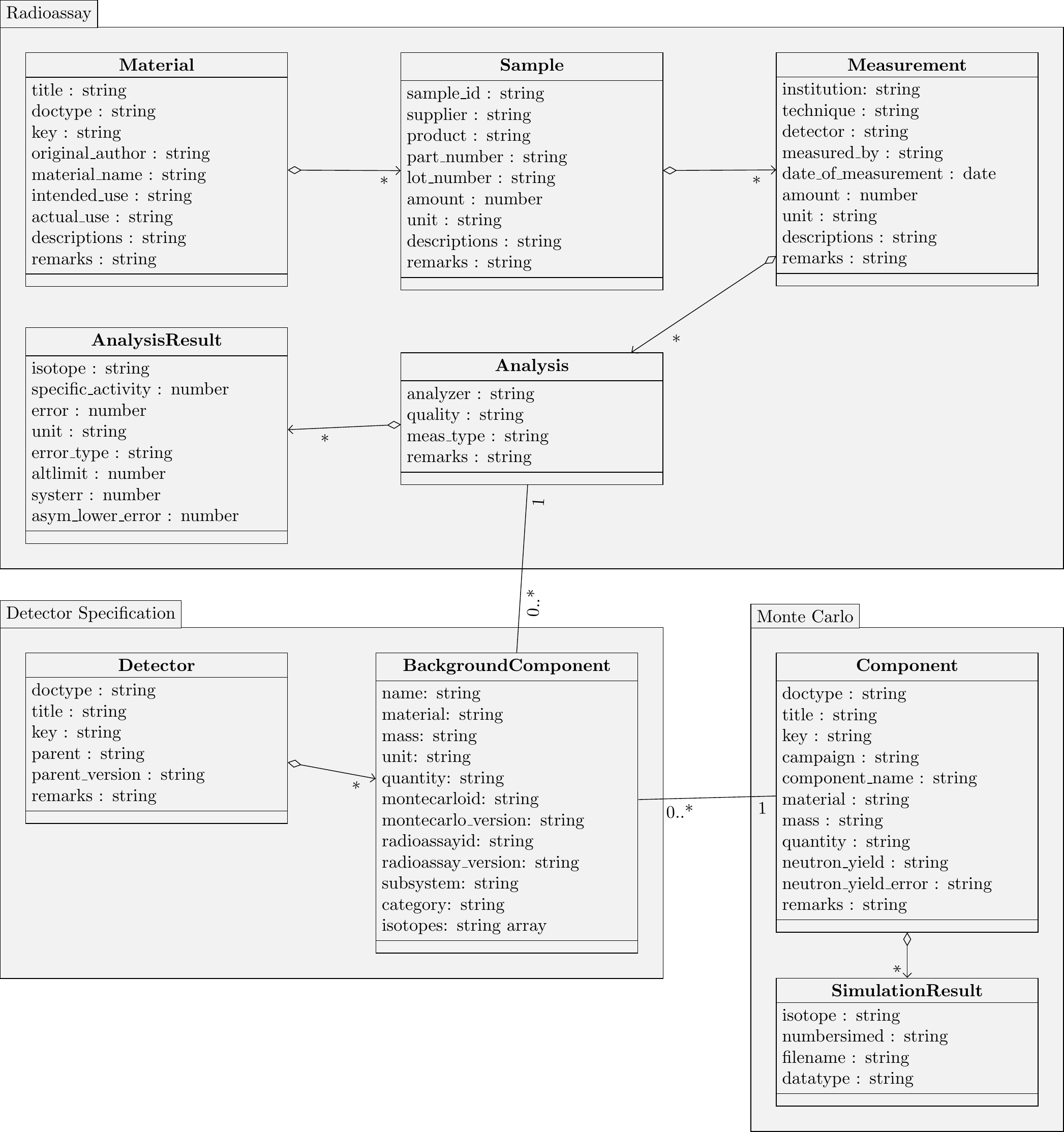} \\
\caption{Schemas of the three most frequently used document types and their mutual associations illustrated in a Unified Modeling Language (UML) diagram. 
         Only a subset of the fields are shown due to limited space.}
\label{fig:schema}
\end{figure*}

Serving as the database on the backend is
a CouchDB instance that is run as a Docker \cite{docker} image. 
The use of a Docker image simplifies future upgrades.
All documents, including \added{\code{Cabinet}} which will be described in the following subsection,
are stored in the main repository of CouchDB. 
Six types of documents have been defined: Radioassay, Monte Carlo, Detector Specification, EXO-200 Radioassay, EXO-200 Radon Emanation, and Assay Requests.
They are distinguished by the ``doctype'' field. 
The schemas of three types of documents are illustrated in Figure~\ref{fig:schema}.

\subsection{User interface component}

The user interface is \added{rendered} by \code{Cabinet}, which is
a collection of JavaScript code stored in a CouchDB repository as a special document called a ``design'' document.
The JavaScript code in \code{Cabinet} is divided into a server-side part, written in CommonJS (a variety of JavaScript intended for server use), and a client-side part written in JavaScript with JQuery.
Running of custom server-side code such as \code{Cabinet} is currently deprecated in version 3 of CouchDB \cite{couchdbv4} and is scheduled to be removed in version 4 \added{\cite{janlehnardt}}.

A typical user query encoded in a URI consists of three parts:
\added{a ``show'', ``list'', or ``search'' function, defined in \code{Cabinet}, that handles the query},
a predefined filter known as a ``view'', and 
the display range for the view.
The HTTP query is first intercepted by CouchDB which generates the requested view.
The list of documents resulting from the view are then passed on to the \added{handling} function.
The \added{handling} function will perform calculations if needed, e.g. unit conversions and calculations of limits, etc.
The result will then be rendered into HTML using the HTML renderer Mustache.
The rendered HTML and the embedded client-side JavaScript code are served to the user's browser.
\code{Cabinet} uses Bootstrap for CSS and JSROOT for histogram display.

With the CouchDB plugin Clouseau \cite{clouseau}, users can perform custom full-text searches in addition to predefined searches.
Clouseau is configured to index selected fields such as author, title, and descriptions of every document.
The indexed keywords can be queried with \added{the ``search''} function
with which the users can perform keyword searches the same way as with a search engine. 
Advanced searches using Lucene query syntax \cite{lucene} are supported \added{by Clouseau}.

Since CouchDB does not have built-in version control, it needs to be provided by our custom code. 
To store previous versions of a document, each document has a ``history'' field which is an array containing copies of previous versions of the document.
When the document is updated, 
the client-side code of \code{Cabinet}
will append the current version of the document, excluding the history field, to the history array.
Currently, 
this version control feature is present only in the client-side code. 
Therefore, when a document is edited through the \added{CouchDB} API,
the version control feature will be bypassed. 
This will be fixed in a future upgrade of the Application.

\code{Cabinet} is equipped with two interactive tools: \code{Allowed Radioactivity Calculator} and \code{Background Investigator}.
Their codes are completely client-side.
This provides direct interactivity in the browser without the need for additional HTTP requests to the server after the initial connection.
Another advantage is that tools are more modular.
Since we anticipate more interactive tools in the future 
for other calculations such as cosmogenics, dust, and radon exposure, 
the high modularity means that such tools can be developed by 
subsystem experts who may not be familiar with the rest of the \code{Cabinet} code.
\added{
These interactive tools are powered by Knockout,
a web framework that separates the display logic and the scientific calculation logic.
The separation of concerns simplifies code maintenance in the future as modifications are more contained.
}

\subsection{Background estimation component: \code{Excelgen}}

The Application provides a comprehensive summary of background contributions of all components
in the form of an Excel spreadsheet that contains all relevant information. 
This is known as the background spreadsheet.
The generation of the background spreadsheet requires access to ROOT and a library for writing an Excel spreadsheet. 
Both of these are not readily available in JavaScript. 
The solution was to develop the code in Python with PyROOT and openpyxl.
The resulting product is called \code{Excelgen}.
\code{Excelgen} can retrieve information, such as radioassay and simulation results, from the CouchDB repositories
and process it into a spreadsheet containing formulas and charts.
It is wrapped in a Flask HTTP server so that \code{Cabinet} can 
initiate the spreadsheet generation through Asynchronous JavaScript and XML (AJAX) calls.
The generated spreadsheet will be saved as a static file to be downloaded by the user through \code{Cabinet}.

\code{Excelgen} consists of 3 parts, a Flask server, a CouchDB API wrapper, and the code that does the main logic.
The Flask server is run as the main program. When it intercepts an AJAX call from \code{Cabinet}, it triggers the main code to start.
The main code calls the CouchDB API wrapper as needed to retrieve necessary information for creating the background spreadsheet.
The background spreadsheets 
include all the formulas necessary to compute the background.
If the user intends to evaluate the impact of different input values, there is no need to download a new spreadsheet.

\subsection{Other interfaces}
The Application needs to exchange information with other databases and software used in nEXO. 
Background allocations for various subsystems are stored in a JIRA database, maintained by Lawrence Livermore National Laboratory (LLNL), for project management purposes. 
These allocations are required to create the last sheet of the background spreadsheet which compares the allocations to the background model at hand.
This is achieved by a standalone Python script that extracts relevant information from the JIRA database, and stores it in a special document in the Database. 
Daily synchronization is deemed sufficient as the update frequency of the allocations is expected to be in the order of once per year.
Another software that connects directly to the Database is the sensitivity calculation code. It retrieves data from the Materials Database using a Python API. 
This interface was essential in the latest sensitivity projection for nEXO \cite{sens2}.
Another interface that will be put in place is the one with an exposure-tracking database which does not exist yet.

\section{Conclusion}
The nEXO Materials Database Application was created primarily to meet the need for a repository of radioassay data.
Over years of development, 
it has evolved into a comprehensive tool that covers the whole 
life cycle of radioassay data generation and consumption --
starting from the initial request for assay measurements, to data storage after measurements, and eventually to the application of the data in various calculations.
In fact, the Application was crucial for the writing of two sensitivity papers that the collaboration has published \cite{sens2, sens1}.
It also facilitated numerous trade studies towards a conceptual design.
\begin{figure*}
\includegraphics[width=\textwidth]{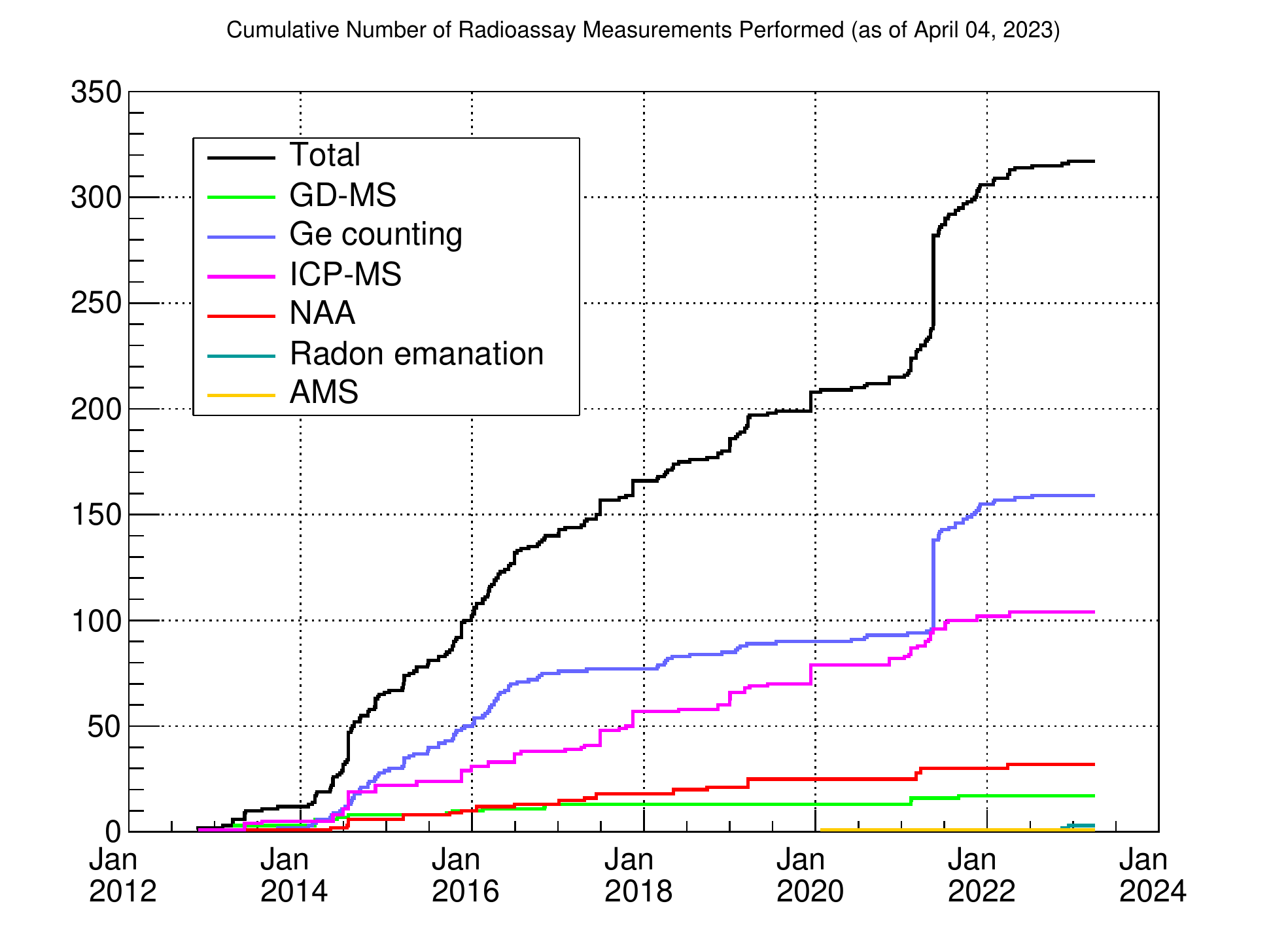} \\
\caption{Time of which radioassay data was entered into the database broken down by assay technique. It also contains entries migrated from the time before Application creation circa 2014.}
\label{fig:throughput}
\end{figure*}
As a testament to its value provided to the experiment,
Figure~\ref{fig:throughput} shows the number of radioassay measurements as a function of time. 
As of Apr 4, 2023, the database stores 317 radioassay measurements, 383 sets of simulation results, and 28 background models.
In addition to these, 316 radioassay measurements and 27 radon measurements from EXO-200's radioassay effort are also stored.

The Materials Database Application has been an indispensable tool for the nEXO project and will continue to support it beyond the design and the construction phases.
In the future, \added{\code{Cabinet} will be replaced by a standalone web application} in anticipation of CouchDB version 4. 
Interfaces to other nEXO databases will be built as needed in the future.

\appendix

\section{External libraries and packages used in the nEXO Materials Database Application}
\label{sec:liblist}
\begin{itemize}
\added{
\item CouchDB (\url{https://couchdb.apache.org/}):
CouchDB is a non-relational document store that stores data as JSON documents.
It also features a HTTP API and a built-in JavaScript interpreter.
These features make CouchApps, web applications hosted on CouchDB, possible.
}
\item JQuery (\url{https://jquery.com/}): 
JQuery is a JavaScript library that simplifies HTML manipulation, AJAX calls, etc.
It is used in the user interface for auto-complete among other interactive features.
\item JSROOT (\url{https://root.cern.ch/js/}):
JSROOT allows the display of ROOT classes such as histograms and graphs in web browsers as interactive elements, similar to what TBrowser does. 
In \code{Cabinet}, it is used to display simulation results.
\added{\item PyROOT (\url{https://root.cern/manual/python/}):
PyROOT is a Python interface for ROOT. It provides access to the full ROOT C++ functionality which is not available from JSROOT.
}
\item Bootstrap (\url{https://getbootstrap.com/}):
Bootstrap is a collection of CSS and JavaScript plugins for the web frontend. It simplifies the graphical design aspect of the user interface development.
It is used throughout \code{Cabinet}.
\item Knockout (\url{https://knockoutjs.com/}):
Knockout is a web framework that implements the Model-View-View Model design pattern. 
It improves code maintainability by separating domain-specific logic, in our case background calculations, from display logic, e.g. what needs to be display, where, and how.
It is used in the two interactive tools in \code{Cabinet}: \code{Allowed Radioactivity Calculator} and \code{Background Investigator}.
\item Flask (\url{https://palletsprojects.com/p/flask/})
Flask is a lightweight web framework written in Python. Its simplicity means that it is ideal for building simple web UIs and APIs. 
In the Application, it is used to build the spreadsheet generator \code{Excelgen} to respond to requests sent from \code{Cabinet}.
Migrating \code{Cabinet} to Flask is the next step in making \code{Cabinet} compatible with CouchDB version 4.
\item openpyxl (\url{https://openpyxl.readthedocs.io})
Openpyxl is a Python library for reading and writing Excel spreadsheets.
This is extensively used in \code{Excelgen} for generating background spreadsheets.
\end{itemize}

\section*{Acknowledgments}

The authors gratefully acknowledge support for this work in part by the DOE Office of Nuclear Physics under grant number DE-FG02-01ER41166.
The nEXO experiment has received support from the Office of Nuclear Physics within DOE's Office of Science, and NSF in the United States; 
from NSERC, CFI, FRQNT, NRC, and the McDonald Institute (CFREF) in Canada; 
from IBS in Korea; from RFBR in Russia; and from CAS and NSFC in China. 
This work was supported in part by Laboratory Directed Research and Development (LDRD) programs at Brookhaven National Laboratory (BNL), 
Lawrence Livermore National Laboratory (LLNL), Oak Ridge National Laboratory (ORNL), Pacific Northwest National Laboratory (PNNL), and SLAC National Accelerator Laboratory.

\bibliographystyle{elsarticle-num} 
\bibliography{matdb}

\begin{thebibliography}{10}
\expandafter\ifx\csname url\endcsname\relax
  \def\url#1{\texttt{#1}}\fi
\expandafter\ifx\csname urlprefix\endcsname\relax\def\urlprefix{URL }\fi
\expandafter\ifx\csname href\endcsname\relax
  \def\href#1#2{#2} \def\path#1{#1}\fi

\bibitem{schechtervalle}
J.~Schechter, J.~W.~F. Valle, Neutrinoless double-beta decay in {SU}(2)
  $\times$ {U}(1) theories, Phys. Rev. D 25 (1982) 2951.

\bibitem{pcdr}
S.~A. Kharusi, et~al.~(\relax{nEXO} Collaboration), {nEXO} pre-conceptual
  design report, {arXiv} (2018) 1805.11142.

\bibitem{exo200}
G.~Anton, et~al.~(\relax{EXO-200} Collaboration), Search for neutrinoless
  double-$\beta$ decay with the complete \relax{EXO-200} dataset, Phys. Rev.
  Lett. 123 (2019) 161802.

\bibitem{sens2}
G.~Adhikari, et~al., {NEXO}: Neutrinoless double beta decay search beyond
  10$^{28}$ year half-life sensitivity, J. Phy. G: Nucl. Part. Phys. 49 (2022)
  015104.

\bibitem{sens1}
J.~Albert, et~al.~(\relax{nEXO} Collaboration), Sensitivity and discovery
  potential of the proposed {nEXO} experiment to neutrinoless double-$\beta$
  decay, Phys. Rev. C 97 (2018) 065503.

\bibitem{geant4a}
J.~Allison, et~al., Recent developments in \relax{Geant4}, Nucl. Instrum. Meth.
  A 835 (2016) 186--225.

\bibitem{geant4b}
J.~Allison, et~al., Geant4 developments and applications, IEEE Trans. Nucl.
  Sci. 53 (2006) 270--278.

\bibitem{geant4c}
S.~Agostinelli, et~al., Geant4 -- a simulation toolkit, Nucl. Instrum. Meth. A
  506 (2003) 250--303.

\bibitem{radsno}
P.~Jagam, J.~Simpson, et~al., Measurements of \relax{Th}, \relax{U} and
  \relax{K} concentrations in a variety of materials, Nucl. Instrum. Meth. A
  324 (1993) 389--398.

\bibitem{radborexino}
C.~Arpesella, et~al.~(\relax{BOREXINO} Collaboration), Measurements of
  extremely low radioactivity levels in \relax{BOREXINO}, Astropart. Phys. 18
  (2002) 1--25.

\bibitem{radlaub}
M.~Laubenstein, et~al., Underground measurements of radioactivity, Appl Radiat.
  Isot. 61 (2004) 167--172.

\bibitem{radexo200a}
D.~Leonard, et~al., Systematic study of trace radioactive impurities in
  candidate construction materials for \relax{EXO-200}, Nucl. Instrum. Meth. A
  591 (2008) 490--509.

\bibitem{radgerda}
D.~Budj\'a\v{s}, et~al., Gamma-ray spectrometry of ultra low levels of
  radioactivity within the material screening program for the \relax{GERDA}
  experiment, Appl Radiat. Isot. 67 (2009) 755--758.

\bibitem{radnext}
V.~\'{A}lvarez, et~al., Radiopurity control in the \relax{NEXT-100} double beta
  decay experiment: procedures and initial measurements, JINST 8 (2013) T01002.

\bibitem{radedelweiss}
E.~Armengaud, et~al., Background studies for the \relax{EDELWEISS} dark matter
  experiment, Astropart. Phys. 47 (2013) 1--9.

\bibitem{radpandax}
X.~Wang, et~al., Material screening with \relax{HPGe} counting station for
  \relax{PandaX} experiment, JINST 11 (2016) T12002.

\bibitem{radmajorana}
N.~Abgrall, et~al., The \relax{Majorana Demonstrator} radioassay program, Nucl.
  Instrum. Meth. A 828 (2016) 22--36.

\bibitem{radxenon1t}
E.~Aprile, et~al., Material radioassay and selection for the \relax{XENON1T}
  dark matter experiment, Eur. Phys. J. C 77 (2017) 890.

\bibitem{radexo200b}
D.~Leonard, et~al., Trace radioactive impurities in final construction
  materials for \relax{EXO-200}, Nucl. Instrum. Meth. A 871 (2017) 169--179.

\bibitem{radlz}
D.~Akerib, et~al., The \relax{LUX-ZEPLIN (LZ)} radioactivity and cleanliness
  control programs, Eur. Phys. J. C 80 (2020) 1044.

\bibitem{radiopurity}
J.~Loach, et~al., A database for storing the results of material radiopurity
  measurements, Nucl. Instrum. Meth. A 839 (2016) 6--11.

\bibitem{bgexp}
B.~Loer, K.~Wierman, bgexplorer: Flask web application to drill down into bill
  of materials and background contributions,
  https://github.com/bloer/bgexplorer, accessed on Mar 21, 2023 (2023).

\bibitem{lrt2017}
D.~Leonard,
  \href{https://indico.ibs.re.kr/event/46/contributions/2750}{Radiopurity
  databases for detector development}, Talk given at Low Radioactivity
  Techniques 2017 in Seoul (2017).
\newline\urlprefix\url{https://indico.ibs.re.kr/event/46/contributions/2750}

\bibitem{couchdb}
\relax{Apache Software Foundation}, Apache
  {CouchDB}\textsuperscript{\textregistered} 3.2.2 documentation,
  https://docs.couchdb.org/en/3.2.2-docs, accessed on Jan 12, 2023 (2023).

\bibitem{rnemanfrac}
A.~Sakoda, Y.~Ishimori, K.~Yamaoka, A comprehensive review of radon emanation
  measurements for mineral, rock, soil, mill tailing and fly ash, App. Rad. and
  Isot. 69 (2011) 1422--1435.

\bibitem{fc}
G.~Feldman, R.~Cousins, Unified approach to the classical statistical analysis
  of small signals, Phys. Rev. D 57 (1998) 3873.

\bibitem{exorad1}
D.~Leonard, et~al., Systematic study of trace radioactive impurities in
  candidate construction materials for {EXO-200}, Nucl. Instrum. Meth. A 591
  (2008) 490--509.

\bibitem{exorad2}
D.~Leonard, et~al., Trace radioactive impurities in final construction
  materials for {EXO-200}, Nucl. Instrum. Meth. A 871 (2017) 169--179.

\bibitem{kapton}
I.~Arnquist, et~al., Ultra-low radioactivity {Kapton} and copper-{Kapton}
  laminates, Nucl. Instrum. Meth. A 959 (2020) 163573.

\bibitem{json}
\relax{ISO/IEC 21778:2017}, {Information technology -- The JSON data
  interchange syntax}, https://www.iso.org/standard/71616.html, accessed on Jan
  12, 2023 (2017).

\bibitem{root}
R.~Brun, F.~Rademakers, {ROOT} - an object oriented data analysis framework,
  Nucl. Instum. Meth. A 389 (1997) 81--86.

\bibitem{docker}
{\color{blue}Merkel, Dirk}, {\color{blue}Docker: lightweight linux containers
  for consistent development and deployment}, Linux journal 2014~(239) (2014)
  2.

\bibitem{couchdbv4}
\relax{Apache Software Foundation}, {CouchDB} documentation: 3. design
  documents, https://docs.couchdb.org/en/stable/ddocs/index.html, accessed on
  Jan 12, 2023 (2023).

\bibitem{janlehnardt}
{\color{blue}Jan Lehnardt}, {\color{blue}Personal communication via {S}lack on
  {M}ay 4, 2023} (2023).

\bibitem{clouseau}
\relax{Cloudant-Labs}, Clouseau: Expose {L}ucene features as an erlang-like
  node, https://github.com/cloudant-labs/clouseau, accessed on Mar 20, 2023
  (2023).

\bibitem{lucene}
\relax{Apache Software Foundation}, Apache {L}ucene, https://lucene.apache.org,
  accessed on Jan 12, 2023 (2023).

\end{thebibliography}


%
%
%
\end{document}